Title: Variational band theory of vibronic polarons in crystals. I. Preamble
Author: M. Georgiev (Institute of Solid State Physics, Bulgarian Academy of Science
   1784 Sofia, Bulgaria)
Comments: 8 pages and 1 figure pdf format
Subj-class: cond-mat

We review the basic theoretical background for working out a variational band solution for vibronic polarons in crystals. It is based on the Lee-Low-Pines proposal as extended by Thomas et al. for describing Jahn-Teller polarons along a linear chain of atoms. The variational properties of antiadiabatic itinerant polarons are also discussed.

1. Introduction

In three parts of a paper we review the development of a variational eigenstate for the vibronic polaron, a specific electron-phonon unit exhibiting both local and itinerant features along a 1D-atomic chain. This Part I recapitulates the general physical principles which should be borne in mind while working out a solution. For instance, the solution should be both translationally invariant, transform according to the irreducible representations of the point group and reflect the mixing character of the electron-phonon coupling. We reproduce some of the math results in the hope that they may be found useful for newcomers. The itinerant behavior of fully blown antiadiabatic polarons from a variational standpoint is also discussed. Part II develops a more refined version based on Merrifield's Variational Ansatz extended so as to cover the system of two electronic bands mixed through coupling to an Einstein phonon of given symmetry. We derive a set of nonlinear variational equations for the phonon and band amplitudes, solved through iterations. The numerical calculations so made are discussed in Part III.

2. Itinerant vibronic polarons

2.1. Lee-Low-Pines solution

The polaron, a solid state entity resulting from the interaction of two fields: fermionic (an electron-hole field) or nearly bosonic (an exciton field) with a boson field (a phonon field), is of interest not only for solid state physics. It may also constitute an example of keen interest for nuclear physics as well, though going far beyond the framework of the present discussion. There is a variety of interaction terms considered in polaron treory, such as the band-diagonal terms coupled to symmetry-retaining vibrational modes which lead to Holstein polarons. At the other extreme are the band-off-diagonal terms coupled to symmetry-breaking modes which lead to vibronic polarons. Consequently, while Holstein's polarons are usually single-band species, vibronic polarons require the availability of more than one fermionic band.

Undoubtedly, it is rather tedious deriving the exact eigenstates of the general vibronic Hamiltonian. In an attempt to mastermind a variational solution appropriate for describing the ground state of a vibronic system, Thomas *et al.* [1,2] have considered a special form of vibronic Hamiltonian which applies to the itinerant Jahn-Teller (JT) effect, the interplay between a moving electron and the local JT distortions:

$$H \equiv H_{el} + H_{latt} + H_{JT} = \varepsilon_0 \sum_l (a_{l1}^\dagger a_{l1} + a_{l2}^\dagger a_{l2}) - \tfrac{1}{2} \sum'_{ll'\gamma} t_{\gamma(ll')} a_{l\gamma}^\dagger a_{l'\gamma} +$$

$$\tfrac{1}{2} \sum_l [\mathbf{P}_l^2 / M + M\omega_s^2 Q_l^2] - \tfrac{1}{2} \sum_{ll'} V_{ll'} Q_l Q_{l'} - G \sum_l Q_l (a_{l1}^\dagger a_{l1} - a_{l2}^\dagger a_{l2}). \tag{1}$$

In the particular case, molecular complexes of tetragonal symmetry are considered whose eigenstates transform according to the E-representation of the tetragonal group $T_d$. The E×β band JT effect mixes two equal-parity electronic bands whose bandgap is vanishing. In the tight-binding approximation, these bands originate from orbital doublets $(\phi_{l\theta}, \phi_{l\varepsilon})$ or $(\phi_{l1}, \phi_{l2})$ of the atomic cluster in a given unit cell l which are two-fold degenerate because of the electron tunneling between clusters in different unit cells. The delocalization caused by tunneling is opposed by the decrease in energy due to the electron-phonon interaction. The resulting instability is relaxed through lowering the symmetry by coupling to the $Q_l$ mode which transforms according to the β-irreducible representation of the tetragonal group. We thus see tunneling delocalization competing with a localization trend due to the JT coupling.

The JT electron-mode coupling term is of the band-diagonal type, proportional to the population difference:

$$H_{JT} = G \sum_l Q_l (n_{l2} - n_{l1}), \tag{2}$$

while the JT electronic energy is proportional to the population sum:

$$H_{el} = \varepsilon_0 \sum_l (n_{l1} + n_{l2}). \tag{3}$$

We are tempted to state outright that the electron-mode coupling in (2) does not effect any mixing of the electronic states of the (two) constituent (degenerate) bands. In contrast, the Pseudo-Jahn-Teller (PJT) coupling $H_{PJT}$ as composed by a band off-diagonal term

$$H_{PJT} = G \sum_l Q_l (a_{l1}^\dagger a_{l2} + a_{l2}^\dagger a_{l1}), \tag{4}$$

represents a genuine mixing for that matter. The PJT electronic energy is proportional to the population difference:

$$H_{el} = \tfrac{1}{2} E_{12} \sum_l (a_{l1}^\dagger a_{l1} - a_{l2}^\dagger a_{l2}) \equiv \tfrac{1}{2} E_{12} \sum_l (n_{l1} - n_{l2}) \tag{5}$$

where $E_{12} = E_2 - E_1 \equiv E_\theta - E_\varepsilon$ is the bandgap (energy reference set at midgap energy).

Either coupling term gives rise to a lattice distortion around the moving electron. For a given distortion pattern $Q_l$, the coupling term is a self-consistent electronic potential moving along with its self-trapped electron. The composite entity is a Jahn-Teller or Pseudo-Jahn-Teller vibronic polaron, respectively. The electron hopping is intraband as interband hopping is prohibited by symmetry.

To evaluate the ground state energy, the following Variational Ansatz is utilized dating back to Lee, Low and Pines:

$$|\phi_{k\gamma}\rangle = C \sum_l exp(i\mathbf{k}\cdot\mathbf{R}_l) \prod_{l'} exp(\alpha_{ll'}(\mathbf{k}\gamma)[b_{l'}^\dagger - b_{l'}]) \sum_{l''} \beta_{ll''}(\mathbf{k}\gamma) a_{l''\gamma}^\dagger |0\rangle \quad (6)$$

where by second quantization of the lattice

$$Q_l = \sqrt{\eta\omega_s / 2M\omega_s^2}\, (b_l^\dagger + b_l)$$

$$P_l = i\sqrt{\eta M\omega_s / 2}\, (b_l^\dagger - b_l) \quad (7)$$

Here $\prod_{l'} exp[\alpha_{ll'}(\mathbf{k}\gamma)(b_{l'}^\dagger - b_{l'})]$ generates a distortion spread at sites l' around a given site l with shape determined by the variational parameters $\alpha_{ll'}(\mathbf{k}\gamma)$ (note that the exponent is proportional to the mode momentum $K_l = P_l / \eta$, $\beta_{ll''}(\mathbf{k}\gamma) a_{l''\gamma}^\dagger$ creates a wavepacket centered at site l where its shape is given by the variational parameters $\beta_{ll''}(\mathbf{k}\gamma)$.

The variational eigenstate $|\phi_{k\gamma}\rangle$ is translationally invariant:

$$T_h|\phi_{k\gamma}\rangle = exp(-i\mathbf{k}\cdot\mathbf{R}_h)|\phi_{k\gamma}\rangle$$

provided the variational parameters depend only on the difference $R_{ll'} = R_l - R_{l'}$, $\alpha_{ll'}(\mathbf{k}\gamma) = \alpha_{l-l''}(\mathbf{k}\gamma)$ and $\beta_{ll''}(\mathbf{k}\gamma) = \beta_{l-l''}(\mathbf{k}\gamma)$. Inversion symmetry also requires an invariance with respect to the change in sign of $R_{ll'}$: $\alpha_{ll'}(\mathbf{k}\gamma) = \alpha_{l'l}(\mathbf{k}\gamma)$ and $\beta_{ll'}(\mathbf{k}\gamma) = \beta_{l'l}(\mathbf{k}\gamma)$.

It will be instructive to reproduce the resulting expressions for the variational energy:

$$\langle\phi_{k\gamma}|H_{el}|\phi_{k\gamma}\rangle = \varepsilon_0\langle\phi_{k\gamma}|\phi_{k\gamma}\rangle - \eta\omega_s C^2 \sum_{l1l2} exp(i\mathbf{k}\cdot[\mathbf{R}_{k1}-\mathbf{R}_{k2}]) \times$$

$$\prod_{l''} exp(-\tfrac{1}{2}[\alpha_{l1-l}^*(\mathbf{k}\gamma)-\alpha_{l2-l}^*(\mathbf{k}\gamma)]^2) \times \sum_{ll'} t_\gamma(ll') \beta_{l1-l}(\mathbf{k}\gamma)\beta_{l2-l'}(\mathbf{k}\gamma), \quad (8)$$

$$\langle\phi_{k\gamma}|H_{latt}|\phi_{k\gamma}\rangle = \tfrac{1}{2}N\eta\omega_s \langle\phi_{k\gamma}|\phi_{k\gamma}\rangle + \eta\omega_s C^2 \sum_{k1k2} exp(i\mathbf{k}\cdot[\mathbf{R}_{k2}-\mathbf{R}_{k1}]) \times$$

$$\prod_{l\sigma} exp(-\tfrac{1}{2}[\alpha_{l1-l\sigma}(\mathbf{k}\gamma)-\alpha_{l2-l\sigma}(\mathbf{k}\gamma)]^2) \sum_{k''} \beta_{l1-l''}(\mathbf{k}\gamma)\beta_{l2-l''}(\mathbf{k}\gamma) \times$$

$$\{\sum_l \alpha_{l1-l}(\mathbf{k}\gamma)\alpha_{l2-l}(\mathbf{k}\gamma) - \tfrac{1}{2}\sum'_{ll'} V_{ll'}(\alpha_{l2-l}(\mathbf{k}\gamma)+\alpha_{l1-l}(\mathbf{k}\gamma))(\alpha_{l2-l'}(\mathbf{k}\gamma)+\alpha_{l1-l'}(\mathbf{k}\gamma))\}, \quad (9)$$

$$\langle\phi_{k\gamma}|H_{JT}|\phi_{k\gamma}\rangle = -\eta\omega_s C^2 \sigma_\gamma G \sum_{l1l2} exp(i\mathbf{k}\cdot[\mathbf{R}_{k2}-\mathbf{R}_{k1}]) \times$$

$$\prod_{l'} exp(-\tfrac{1}{2}[\alpha_{l1-l'}(\mathbf{k}\gamma)-\alpha_{l2-l'}(\mathbf{k}\gamma)]^2) \sum_l (\alpha_{l1-l}(\mathbf{k}\gamma)+\alpha_{l2-l}(\mathbf{k}\gamma)) \beta_{l1-l}(\mathbf{k}\gamma)\beta_{l2-l}(\mathbf{k}\gamma), \quad (10)$$

with the normalization constant

$$1 = \langle\phi_{k\gamma}|\phi_{k\gamma}\rangle = C^2 \sum_{l1l2} exp(i\mathbf{k}\cdot[\mathbf{R}_{k2}-\mathbf{R}_{k1}]) \times$$

$$\prod_{l'} exp(-\tfrac{1}{2}[\alpha_{l1-l'}(\mathbf{k}\gamma)-\alpha_{l2-l'}(\mathbf{k}\gamma)]^2) \sum_{l''} \beta_{l1-l''}(\mathbf{k}\gamma)\beta_{l2-l''}(\mathbf{k}\gamma) \quad (11)$$

and the following notations used:

$t_\gamma(ll') = t_\gamma(ll') / (\eta\omega_s)$, $\eta\omega_s V_{ll'} = V_{ll'} / (2M\omega_s^2)$, $\eta\omega_s G = G\sqrt{(\eta\omega_s/2M\omega_s^2)}$,

with $G^2 = E_{JT} / \eta\omega_s$ because $E_{JT} = G^2/2K$, $K = M\omega_s^2$; $\sigma_\gamma = 1$ ($\gamma=1$), $\sigma_\gamma = -1$ ($\gamma=2$). Finally the functional $E_\gamma = <\phi_{k\gamma}|H|\phi_{k\gamma}> / <\phi_{k\gamma}|\phi_{k\gamma}>$ is minimized with respect to each of the two sets of parameters $\alpha_l(k\gamma)$ and $\beta_l(k\gamma)$.

The PJT case is dealt with similarly. Differences with the JT Hamiltonian appear in both the electronic and mixing terms. We arrive at the following PJT energy components:

$$<\phi_{k\gamma}|H_{el}|\phi_{k\gamma}> = \tfrac{1}{2}E_{12}<\phi_{k\gamma}|\Sigma(n_1-n_2)|\phi_{k\gamma}> - \eta\omega_s C^2 \Sigma_{l1l2}\, exp(i\mathbf{k}.[\mathbf{R}_{k1}-\mathbf{R}_{k2}])\times$$

$$\prod_{l''} exp(-\tfrac{1}{2}[\alpha_{l1-l}*(k\gamma) - \alpha_{l2-l}*(k\gamma)]^2) \sum_{ll'} t_\gamma(ll')\beta_{l1-l}(k\gamma)\beta_{l2-l'}(k\gamma), \quad (12)$$

with

$$\tfrac{1}{2}E_{12}<\phi_{k\gamma}|\Sigma(n_1-n_2)|\phi_{k\gamma}> = \tfrac{1}{2}E_{12}(n_1-n_2)<\phi_{k\gamma}|\phi_{k\gamma}> = \tfrac{1}{2}E_{12}\sigma_\gamma<\phi_{k\gamma}|\phi_{k\gamma}>, \quad n_1+n_2=1,$$

$$<\phi_{k\gamma}|H_{latt}|\phi_{k\gamma}> = \tfrac{1}{2}N\eta\omega_s<\phi_{k\gamma}|\phi_{k\gamma}> + \eta\omega_s C^2 \Sigma_{k1k2}\, exp(i\mathbf{k}.[\mathbf{R}_{k2}-\mathbf{R}_{k1}])\times$$

$$\prod_{l\sigma} exp(-\tfrac{1}{2}[\alpha_{l1-l\sigma}(k\gamma) - \alpha_{l2-l\sigma}(k\gamma)]^2) \sum_{k''} \beta_{l1-l''}(k\gamma)\beta_{l2-l''}(k\gamma)\times$$

$$\{\Sigma_l \alpha_{l1-l}(k\gamma)\, \alpha_{l2-l}(k\gamma) - \tfrac{1}{2}\Sigma'_{ll'}V_{ll'}(\alpha_{l2-l}(k\gamma)+\alpha_{l1-l}(k\gamma))(\alpha_{l2-l'}(k\gamma)+\alpha_{l1-l'}(k\gamma))\} \quad (13)$$

$$<\phi_{k\gamma}|H_{PJT}|\phi_{k\gamma}> = -\eta\omega_s C^2 G \Sigma_{l1l2} exp(i\mathbf{k}.[\mathbf{R}_{k2}-\mathbf{R}_{k1}])\prod_{l'} exp(-\tfrac{1}{2}[\alpha_{l1-l'}(k\gamma)-\alpha_{l2-l'}(k\gamma)]^2)\times$$

$$\Sigma_l (\alpha_{l1-l}(k\gamma) + \alpha_{l2-l}(k\gamma))\beta_{l1-l}(k\gamma)\beta_{l2-l}(k\gamma), \quad (14)$$

with the normalization constant as above.

There have been a number of proposals for a variational Ansatz found useful for various applications. Among these is Merrifield's Ansatz to be discussed at length below. At this point we reproduce Merrifield's variational eigenstate:[3]

$$|\psi_\mu(\kappa)> = N^{-1/2} \Sigma_n exp(i\kappa n)\, a_{n\mu}^\dagger\, F_{n\mu}^\kappa |0>$$

$$F_{\mu n}^\kappa = exp\{-N^{-1/2} \Sigma_q [\beta_{q\mu}^\kappa exp(-iqn) b_q^\dagger - \beta_{q\mu}^{\kappa*} exp(+iqn) b_q]\}$$

where $\kappa$ is the total crystalline momentum, $q$ is the phonon momentum and $\mu$ is the electronic band label. $\beta_{q\mu}^\kappa$ are the phonon amplitudes, $a_{n\mu}$ and $b_q$ are the fermion and boson ladder operators, respectively. Merrifield's eigenstate is a superposition of plane waves modulated by the phonon form-factors $F_{\mu n}^\kappa$. Merrifield's Ansatz has originally been aimed at applying to Holstein's polaron.

Calculated phonon amplitudes $\beta_{q\mu}^\kappa$ of antiadiabatic small polarons ($G_{\mu\mu}^2 > 2J_\mu$) are shown in Figure 1 as obtained by solving the variational equations by way of iterations. These are to be compared with corresponding mappings elsewhere.[4] We see both small polaron (slow varying with κ) and large polaron (fast varying with κ) features, though the latter ones are predominating.

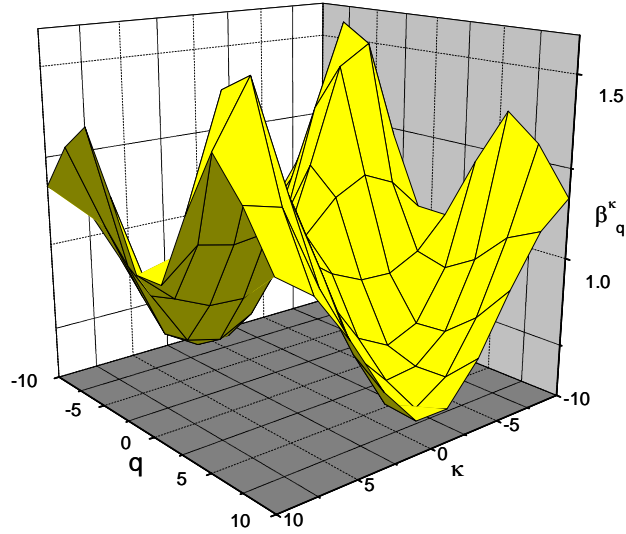

.

Figure 1

Two-dimensional mappings in of the phonon amplitudes $\beta_{q\mu}^{\kappa}$ of Merrifield-Holstein's antiadiabatic polarons in the (q,κ) space of phonon momentum q and total crystalline momentum κ calculated at $J_\mu = 0.2$ (electron hopping energy) and $G_{\mu\mu} = 1$ (electron-phonon coupling constant). Antiadiabatic polarons form when $G_{\mu\mu}^2 > 2J_\mu$.

2.2. Antiadiabatic small polarons

The adiabaticity criterion in itinerancy,[5] as formulated in temporal terms, is that the electronic motion across the lattice is antiadiabatic if the extra charge stays at site l for a time $\tau_l$ that is long compared with the period of the symmetry-breaking lattice vibration $\tau_{vib}$: $\tau_l \gg \tau_{vib} = 2\pi/\omega_s$. In this case the local distortion has enough time to develop fully. In as much as time and energy are reciprocal in quantum theory, we rewrite the above criterion to read t « ηω$_s$ « E$_{JT}$, where t is the intersite hopping term and E$_{JT}$ is the Jahn-Teller energy. In this case the associated lattice distortion is complete and follows the motion of the electron without retardation. On the contrary the electron motion will be adiabatic if too fast to let lattice distortions grow. Now, the electron hopping energy exceeds the local phonon quantum t » ηω$_s$, though not necessarily the local JT energy as well. It is clear that the antiadiabatic polarons are closer to the idealized concept of a charge carrier coupled to its self-erected lattice distortion.

For a fully developed JT polaron at $E_{JT} \gg \eta\omega_s > t$, the distortion is squeezed, $\alpha_{ll'}(\mathbf{k}\gamma) = \alpha_{ll}(\mathbf{k}\gamma)\delta_{ll'}$, and the wave packet is narrow, $\beta_{ll''}(\mathbf{k}\gamma) = \beta_{ll}(\mathbf{k}\gamma)\delta_{ll''}$, because $\sum_{l''}\beta_{ll''}(\mathbf{k}\gamma)\alpha_{l''\gamma}^\dagger|0> = \beta_{ll}(\mathbf{k}\gamma)\alpha_{l\gamma}^\dagger|0>$; the polaron is localized at site l in a local distortion pattern:

$$|\phi_{\mathbf{k}\gamma}> = C\sum_l exp(i\mathbf{k}.\mathbf{R}_l)\, exp(\alpha_{ll}(\mathbf{k}\gamma)\,[b_l^\dagger - b_l])\, \beta_{ll}(\mathbf{k}\gamma)\, a_{l\gamma}^\dagger|0> \qquad (15)$$

This is the case of strong JT coupling or small JT polaron. For small displacements $\alpha_{ll'}(\mathbf{k}\gamma)$ the state $|\phi_{\mathbf{k}\gamma}>$ is more spread out which is the case of weak JT coupling or large JT polaron. For constant $\beta_{ll''}(\mathbf{k}\gamma) = \beta$, we get the limit of an unbound electron-phonon pair.

It would be useful to rewrite the variational energy constituents of a small polaron:

$$<\phi_{\mathbf{k}\gamma}|H_{el}|\phi_{\mathbf{k}\gamma}> = \varepsilon_0<\phi_{\mathbf{k}\gamma}|\phi_{\mathbf{k}\gamma}> - \eta\omega_s C^2 \sum_{l1l2} exp(i\mathbf{k}.[\mathbf{R}_{k1}-\mathbf{R}_{k2}])\times$$

$$exp(-\tfrac{1}{2}[\alpha_{l1l1}(\mathbf{k}\gamma)-\alpha_{l2l2}(\mathbf{k}\gamma)]^2)\, t_{\gamma(l1l2)}\beta_{l1l1}(\mathbf{k}\gamma)\beta_{l2l2}(\mathbf{k}\gamma)$$

$$= \varepsilon_0<\phi_{\mathbf{k}\gamma}|\phi_{\mathbf{k}\gamma}> - \eta\omega_s C^2 \sum_{l1l2} exp(i\mathbf{k}.[\mathbf{R}_{k1}-\mathbf{R}_{k2}])\, t_{\gamma(l1l2)}\beta(\mathbf{k}\gamma)^2 \qquad (16)$$

$$<\phi_{\mathbf{k}\gamma}|H_{latt}|\phi_{\mathbf{k}\gamma}> = \tfrac{1}{2}N\eta\omega_s<\phi_{\mathbf{k}\gamma}|\phi_{\mathbf{k}\gamma}> + \eta\omega_s C^2 \sum_{k1k2} exp(i\mathbf{k}.[\mathbf{R}_{k2}-\mathbf{R}_{k1}])\times$$

$$exp(-\tfrac{1}{2}[\alpha_{l1l1}(\mathbf{k}\gamma)-\alpha_{l2l2}(\mathbf{k}\gamma)]^2)\alpha_{l1l1}(\mathbf{k}\gamma)\alpha_{l2l2}(\mathbf{k}\gamma)\beta_{l1l1}(\mathbf{k}\gamma)\beta_{l2l2}(\mathbf{k}\gamma)$$

$$= \tfrac{1}{2}N\eta\omega_s<\phi_{\mathbf{k}\gamma}|\phi_{\mathbf{k}\gamma}> + \eta\omega_s C^2\sum_{k1k2} exp(i\mathbf{k}.[\mathbf{R}_{k2}-\mathbf{R}_{k1}])\alpha(\mathbf{k}\gamma)^2\beta(\mathbf{k}\gamma)^2 \qquad (17)$$

$$<\phi_{\mathbf{k}\gamma}|H_{JT}|\phi_{\mathbf{k}\gamma}> = -\eta\omega_s C^2 \sigma_\gamma G \sum_{l1l2} exp(i\mathbf{k}.[\mathbf{R}_{k2}-\mathbf{R}_{k1}])\, exp(-\tfrac{1}{2}[\alpha_{l1l1}(\mathbf{k}\gamma)-\alpha_{l2l2}(\mathbf{k}\gamma)]^2)\times$$

$$(\alpha_{l1l1}(\mathbf{k}\gamma)+\alpha_{l2l2}(\mathbf{k}\gamma))\beta_{l1l1}(\mathbf{k}\varsigma)\beta_{l2l2}(\mathbf{k}\gamma)$$

$$= -2\eta\omega_s C^2 \sigma_\gamma G \sum_{l1l2} exp(i\mathbf{k}.[\mathbf{R}_{k2}-\mathbf{R}_{k1}])\alpha(\mathbf{k}\gamma)\beta(\mathbf{k}\gamma)^2 \qquad (18)$$

with the normalization constant obtained from

$$1 = <\phi_{\mathbf{k}\gamma}|\phi_{\mathbf{k}\gamma}> \equiv C^2\sum_{l1l2} exp(i\mathbf{k}.[\mathbf{R}_{k2}-\mathbf{R}_{k1}])exp(-\tfrac{1}{2}[\alpha_{l1l1}(\mathbf{k}\gamma)-\alpha_{l2l2}(\mathbf{k}\gamma)]^2)\beta_{l1l1}(\mathbf{k}\gamma)\beta_{l2l2}(\mathbf{k}\gamma)$$

$$= C^2\sum_{l1l2} exp(i\mathbf{k}.[\mathbf{R}_{k2}-\mathbf{R}_{k1}])\beta(\mathbf{k}\gamma)^2 \qquad (19)$$

where we neglect for simplicity the vibrational dispersion $V_{ll'} = 0$ and set $\alpha_{ll'} = \alpha$, $\beta_{ll'} = \beta$. We further minimize by way of:

$$\delta\{<\phi_{\mathbf{k}\gamma}|H|\phi_{\mathbf{k}\gamma}>/<\phi_{\mathbf{k}\gamma}|\phi_{\mathbf{k}\gamma}>\} \equiv \delta\{\varepsilon_0 - t_{\gamma(ll')} + \tfrac{1}{2}N\eta\omega_s + \eta\omega_s\alpha(\mathbf{k}\gamma)^2 - 2\eta\omega_s\sigma_\gamma G\alpha(\mathbf{k}\gamma)\}$$

$$= \eta\omega_s\delta\{\alpha(\mathbf{k}\gamma)^2 - 2\sigma_\gamma G\alpha(\mathbf{k}\gamma)\} = 2\eta\omega_s[\alpha(\mathbf{k}\gamma)-\sigma_\gamma G]\delta\alpha(\mathbf{k}\gamma) = 0 \qquad (20)$$

which yields

$$\alpha(\mathbf{k}\gamma) = \sigma_\gamma G = \sigma_\gamma\sqrt{(E_{JT}/\eta\omega_s)} \qquad (21)$$

whereas we can set $\beta(\mathbf{k}\gamma) \equiv 1$. Using the result the wavepacket is:

$$|\phi_{\mathbf{k}\gamma}\rangle = C \sum_l exp(i\mathbf{k}.\mathbf{R}_l) \, exp\{\sigma_\gamma\sqrt{(E_{JT}/\eta\omega_s)}[\, b_l^\dagger - b_l]\} a_{l\gamma}^\dagger |0\rangle$$

$$= C \sum_l exp(i\mathbf{k}.\mathbf{R}_l) \, exp(-i\sigma_\gamma K_l L) \, a_{l\gamma}^\dagger |0\rangle \qquad (22)$$

with

$$C = \sum_{l1 l2} exp(i\mathbf{k}.[\mathbf{R}_{k2}-\mathbf{R}_{k1}])^{-\frac{1}{2}} \qquad (23)$$

Here $L = 2\sqrt{(\eta^2 E_{JT}/2M)}/(\eta\omega_s) = G/M\omega_s^2 = G/K$ is the characteristic distortion length, *viz.* the small JT-polaron radius.

The ground-state energy is dependent on the band label:

$$E_{g\gamma} = \varepsilon_0 - t_{\gamma(ll')} + \tfrac{1}{2}N\eta\omega_s + \eta\omega_s\alpha(\mathbf{k}\gamma)^2 - 2\eta\omega_s\sigma_\gamma G\alpha(\mathbf{k}\gamma)$$

$$= \varepsilon_0 - t_{\gamma(ll')} - E_{JT} + \tfrac{1}{2}N\eta\omega_s \qquad (24)$$

via the hopping term. Both hopping and JT localization are seen to stabilize the small polaron. In so far as the ground state energy $E_{g\gamma}$ is independent of the electron crystalline momentum $\mathbf{k}$, the small JT-polaron is ultimately localized in the extreme limit under consideration.

The variational principle applies likewise to the PJT polaron as well, leading to

$$\alpha(\mathbf{k}\gamma) = G = \sqrt{(E_{JT}/\eta\omega_s)}, \qquad (25)$$

while its ground-state energy obtains as

$$E_{g\gamma} = \tfrac{1}{2}E_{12}\sigma_\gamma - t_{\gamma(ll')} + \tfrac{1}{2}N\eta\omega_s + \eta\omega_s\alpha(\mathbf{k}\gamma)^2 - 2\eta\omega_s G\alpha(\mathbf{k}\gamma)$$

$$= \tfrac{1}{2}E_{12}\sigma_\gamma - t_{\gamma(ll')} - E_{JT} + \tfrac{1}{2}N\eta\omega_s \qquad (26)$$

## 3. Conclusion

We reviewed the primary steps in working out a variational solution aimed at describing the vibronic polaron unit manifesting both local and itinerant features. However, the advantages of the variational Ansatz have only been studied for a linear atomic chain. Immediate improvements should involve an extension to square 2D-lattices which merit special attention to model conducting Cu-O planes in high-$T_c$ superconducting materials.